\documentclass[twocolumn,twoside,slac]{revtex4}
\usepackage{graphicx}
\usepackage{fancyhdr}
\begin{document}

\title{GraXML - Modular Geometric Modeler}

\author{Julius H\v{r}ivn\'{a}\v{c}}
\affiliation{LAL, Orsay, France}

\begin{abstract}
Many entities managed by HEP Software Frameworks represent spatial 
(3-dimensional) real objects. Effective definition, manipulation and 
visualization of such objects is an indispensable functionality.

GraXML is a modular Geometric Modeling toolkit capable of processing geometric
data of various kinds (detector geometry, event geometry) from different
sources and delivering them in ways suitable for further use. Geometric data 
are first modeled in one of the Generic Models. Those Models are then used to 
populate powerful Geometric Model based on the Java3D technology. While Java3D 
has been originally created just to provide visualization of 3D objects, its 
light weight and high functionality allow an effective reuse as a general 
geometric component. This is possible also thanks to a large overlap between 
graphical and general geometric functionality and modular design of Java3D 
itself. Its graphical functionalities also allow a natural visualization of 
all manipulated elements.

All these techniques have been developed primarily (or only) for the Java 
environment. It is, however, possible to interface them transparently to 
Frameworks built in other languages, like for example C++.

The GraXML toolkit has been tested with data from 
several sources, as for example ATLAS and ALICE detector description and ATLAS 
event data. Prototypes for other sources, like Geometry Description Markup 
Language (GDML) exist too and interface to any other source is easy to add. 
\end{abstract}
\maketitle

\thispagestyle{fancy}

\section{History}

GraXML~\cite{GraXML},\cite{GraXML-talk},\cite{JAGDD} has been originally developed as 
a simple 3D Detector Display for
ATLAS Generic Detector Description (AGDD)~\cite{AGDD}, i.e. for the detector geometry
specified in a generic (application neutral) way using XML. (Since then, many
other experiments and Geant4 have followed ATLAS example.)

Later, the support for several other geometry descriptions has been added to GraXML:
\begin{itemize}
\item {\bf AliDD} is AGDD extended with geometric elements used in ALICE, but
  not in ATLAS.
\item {\bf AGDD v6} is AGDD extended with mathematical formulas to express 
  complex relations between elements and with an access to a relational database to 
  get initial parameters.
\item {\bf AtlasEvent}~\cite{AtlasEvent} is a format used to describe ATLAS Event data, like
  hits and tracks.
\item {\bf AtlantisEvent} (prototype) is similar to AtlasEvent, it is used by
  the Atlantis~\cite{Atlantis} event display.
\item {\bf GDML}~\cite{GDML} (prototype) is an XML format of Geant4 detector description.
\item Direct access via API is available from Java directly and from C++ via
  proxies generated by JACE~\cite{JACE}.
\end{itemize}

Quite soon, it has been realized that requirements on Event and Detector Display
functionality often closely correspond to requirements on Geometric Modeler:
\begin{itemize}
\item Both have to provide a logical navigation in complex geometric structures.
\item Both have to provide a complex geometrical navigation functionality, like:
  \begin{itemize}
  \item identification of the element occupying a spacial position
    (``Where am I ?''), 
  \item intersections between elements (picking, collision detection),
  \item attaching of materials to elements,
  \item displacement and calibration (interactivity).
  \end{itemize}
\item Both need highly optimized geometry description in memory, supporting
  millions of 3D objects.
\end{itemize}
The only functionality which is special in an Display application is the manipulation of
visual object's properties, like transparency, shading, etc. This 
functionality has, however, very big demand on system resources 
(mainly memory).

It has been concluded, that 3D Geometric Engine can be used as Geometric Modeler
foundation if its visual overhead can be removed. Java3D, which is used as 
a Geometric Engine in GraXML, can be used without its visual overhead so
it can serve as a general Modeler foundation.

GraXML has been then re-engineered to provide flexible Geometric Modeling
Toolkit with optional Display capabilities. Currently, it contains following
components:
\begin{itemize}
\item {\bf Core} provides the foundation for the toolkit, it uses Java3D
  library for a description of 3D objects.
\item {\bf Generic Modelers} are used to parse XML geometric descriptions
  and create their memory model.
\item {\bf Geometric Modelers} create optimized 3D model of the geometry.
\item {\bf Applications} (visual or not) are small programs build on top
  of the foundation toolkit.
\item {\bf Utilities} help to manage and organize data.
\end{itemize} 

\begin{figure}[!]
\centering
\includegraphics[width=80mm]{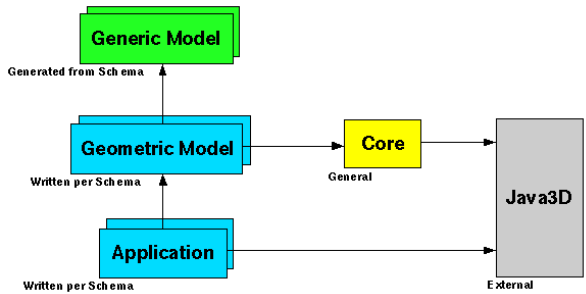}
\caption{Overview of the GraXML Toolkit Global Architecture.} 
\label{History}
\end{figure}

\section{Architecture}

\begin{figure*}[!]
\centering
\includegraphics[width=135mm]{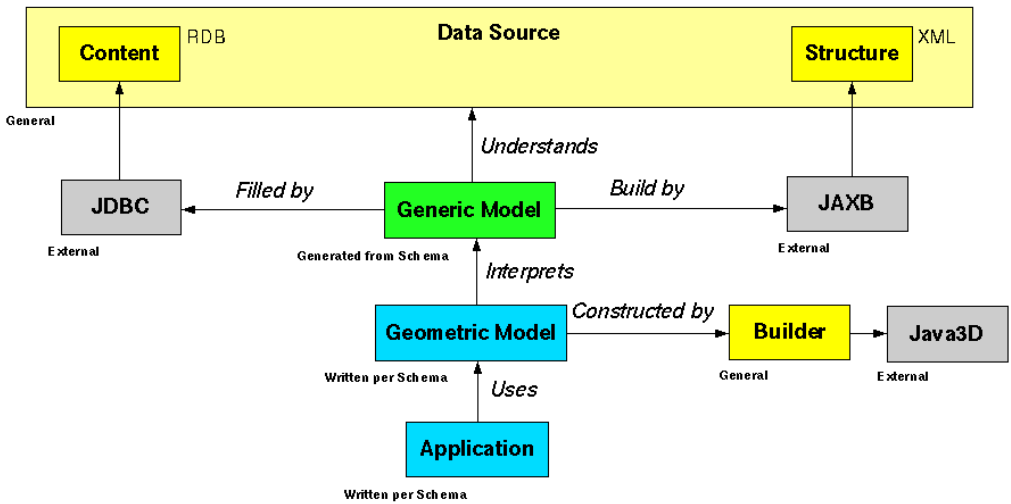}
\caption{GraXML Toolkit Global Architecture.} 
\label{Architecture}
\end{figure*}

\subsection{Generic Model}

\begin{figure*}[!]
\centering
\includegraphics[width=135mm]{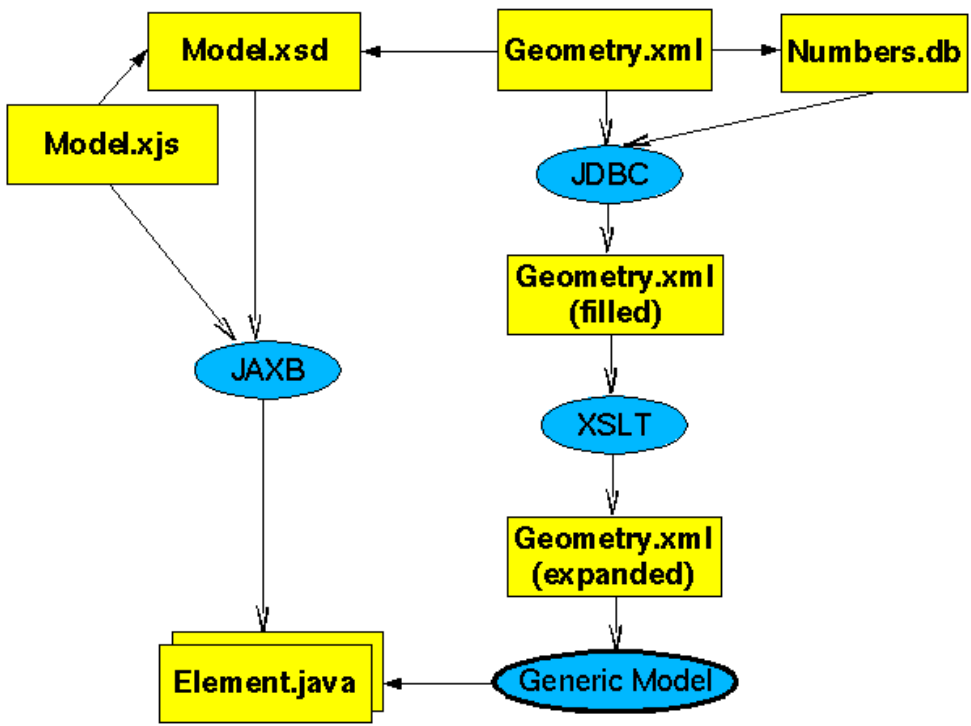}
\caption{Generic Model Architecture.} 
\label{Generic}
\end{figure*}

A special Generic Model is created for each source data XML Schema definition. 
JAXB~\cite{JAXB} processor is used to create an object model similar to widely
known DOM model, but with several advantages. JAXB model is tuned to the XML
Schema which has been used to create it. The main features of the JAXB model are:
\begin{itemize}
\item Created classes correspond to available XML elements. 
\item Their methods (setters/getters) correspond directly to elements 
  attributes.
\item Variables have proper types and default values based on a Schema definition
  and an optional customization file.
\item The generation can be customized (via an XML customization file) to capture more
  complex structures and relations.
\item Created classes can be modified by extension or by helper conversion
  classes.
\end{itemize}
The generated object model is more natural and faster than traditionally used DOM and
SAX-based models. Listing~\ref{JAXB} shows a simple example of a class corresponding
to an XML element.

\begin{table}[!]
\begin{center}
\begin{tabular}{|l|}
\hline
$<box\ x="1.1"\ y="2.2"\ z="3.3"/>$ \\
\hline
$public\ class\ box\ \{$\\
$\ \ public\ double\ x;$\\
$\ \ public\ double\ y;$\\
$\ \ public\ double\ z;$\\
$\ \ \}$ \\
\hline
\end{tabular}
\caption{Simplified example of an XML element and corresponding class
  generated by JAXB processor.}
\label{JAXB}
\end{center}
\end{table}

\subsubsection{Initial values}

A detector description XML file can have its initial values outsourced into a
Relational Database. The Nova~\cite{NOVA} database is used for AGDD. XSQL~\cite{XSQL} Schema is used to
define connections between the XML file and the Relational Database. A simple JDBC~\cite{JDBC}
Connection is then used to fill those values into a parametrised XML file as is demonstrated in
the Listing~\ref{XSQL}.

\begin{table}[!]
\begin{center}
\begin{tabular}{|l|}
\hline
$<XSQLConfig>$ \\
$<connectiondefs>$ \\
$\ \ <connection\ name="demo">$ \\
$\ \ \ \ \dots$ \\
$\ \ \ \ <dburl>jdbc:mysql://nova.site.org/NOVA</dburl>$ \\
$\ \ \ \ </connection>$ \\
$\ \ </connectiondefs>$ \\
$\ \ </XSQLConfig>$ \\
$\dots$ \\
$<var\ connection="demo"\ name="SCT.length"/>$ \\
$\dots$ \\
\hline
$<var\ name="SCT.length"\ value="123.456"/>$ \\
\hline
\end{tabular}
\caption{Definition of the connetion to a relational database and the result 
  of the filling operation in AGDD XML v.6.}
\label{XSQL}
\end{center}
\end{table}

\subsubsection{Formulas}

Symmetries and dependencies between attributes in XML file can be expressed 
using standard mathematical formulas, like in the Listing~\ref{Formulas}. 
The XSLT~\cite{XSLT} stylesheets with the simple Java evaluator
are then used to expand structures with formulas to concrete elements. Formulas
together with initial values read in from a relational database help to make the XML
files very compact.

\begin{table}[!]
\begin{center}
\begin{tabular}{|l|}
\hline
$<array\ name="a"\ values="1;2;3;4;5;6;7;8;9;10"/>$ \\
$<table\ name="t">$ \\
$\ \ <row\ values="1;2;3;4;5"/>$ \\
$\ \ <row\ values="6;7;8;9;10"/>$ \\
$\ \ </table>$ \\
$<var\ name="a0"\ value="1"/>$ \\
$<var\ name="b"\ value="a0*2"/>$ \\
$<var\ name="c"\ value="a[2]*a[3]"/>$ \\
$\dots$ \\
$<box\ XYZ="5.5;a[5];t[2,3]"\ name="abox"/>$ \\
\hline
\end{tabular}
\caption{Examples of formulas in AGDD XML v.6.}
\label{Formulas}
\end{center}
\end{table}

\subsection{Geometric Model}

The Java3D~\cite{Java3D} SceneGraph is build from the Generic Model according to 
several Build Options:
\begin{itemize}
\item Presence of {\bf Graphical} visual attributes and
  functionality.
\item {\bf Level of Optimization} specifies how much will be created SceneGraph optimized.
  This mainly influences the strategy employed to share
  representations (SharedGroups) for identical or similar structures. Repeated
  structures are discovered during SceneGraph building and reused as
  SharedGroups. Higher
  Levels of Optimization allow creation of very compact Geometric Models. The
  interactivity (and calibration capabilities) of such optimized models are
  limited as many volumes don't exist as individual entities and can't then be
  individually manipulated. A more optimized SceneGraph is smaller and faster
  but allows only limited interactivity or calibration.
\item {\bf Level of Quality} influences how closely are 3D objects represented.
  Its consequences are for example the level of approximation of curved surfaces or
  using of visual enhancements like anti-aliasing. The effects of Level of Quality
  are restricted by a chosen Level of Optimization. 
\item {\bf Level of Interactivity} specifies how interactive Display will be
  and how much objects can be calibrated. A higher Level of Interactivity allows
  higher possibilities to change objects properties at run-time. The effects of 
  Level of Interactivity are restricted by the chosen Level of Optimization and
  Quality. Depending on the Level of Optimization, SceneGraph elements
  (Shapes, Groups) can be modified at run time to provide a dynamic functionality
  like:
  \begin{itemize}
  \item {\bf Calibration}
  \item {\bf Graphics Operations}:
    \begin{itemize}
    \item Modification of {\bf Real Objects} (their shape, place, orientation, 
      \dots),
    \item Modification of the Objects {\bf Visual Characteristics} (visibility, color,
      transparency,\dots).
    \end{itemize}
  \end{itemize}
\item {\bf Representations} to be used to represent Generic Objects and their 
  properties. 
\end{itemize}

\begin{figure}[!]
\centering
\includegraphics[width=80mm]{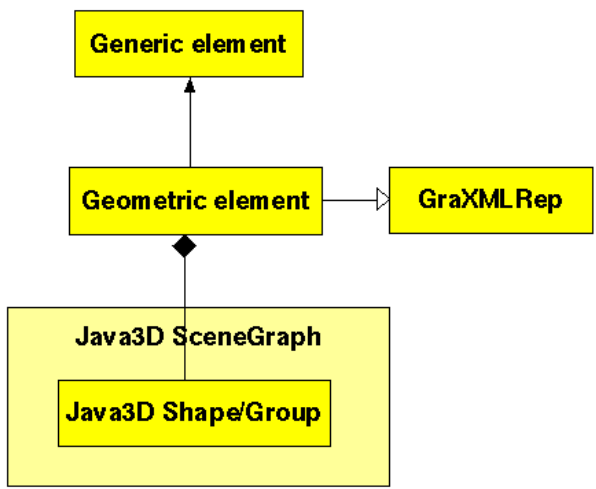}
\caption{Geometric Model Architecture.} 
\label{Geometric}
\end{figure}

\begin{figure*}[!]
\centering
\includegraphics[width=135mm]{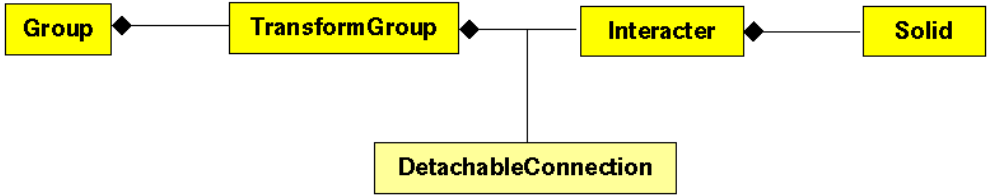}
\caption{SceneGraph extensions.} 
\label{SceneGraph}
\end{figure*}

Additional Geometric Model functionalities are generally added as special nodes inserted
into a SceneGraph as is described in Figure~\ref{SceneGraph}.

As SceneGraphs (or their subgraphs) are compiled for speed, Builder Options
can't be changed later (when Scene Graph is active and used).

All Geant4 CSG Solids (and some others) have been implemented as standard
Java3D Shapes with constructors equivalent to the Geant4 constructors. 
Those solids are usually specializations of more generic Shapes.

New Shapes (Helix, \dots) or special Shapes (Outline, \dots) are also included.

All those Shapes have been contributed to the FreeHEP~\cite{FreeHEP} library.

\section{Data Sources}

GraXML can get data in various ways, preferably in the XML format. The system is
easily extendable. GraXML currently supports following data sources:
\begin{itemize}
\item {\bf XML}  
  \begin{itemize}
  \item {\bf Detector Description}
    \begin{itemize}
    \item {\bf AGDD v4}~\cite{AGDD} is the original explicit ATLAS Generic Detector
      Description.
    \item {\bf AliDD} is AGDD with additional elements, used in ALICE, but not in 
      ATLAS.
    \item {\bf AGDD v6} is AGDD with arithmetic formulas and connection to
      a relational database.
    \item {\bf GDML}~\cite{GDML} (prototype) is the Geant4 format of geometric description.
    \end{itemize}
  \item {\bf Events}
    \begin{itemize}
    \item {\bf AtlasEvent}~\cite{AtlasEvent} is the XML format of ATLAS Event data.
    \item {\bf AtlantisEvent} is the XML format of ATLAS Event data used by
      Atlantis~\cite{Atlantis} Event Display.
    \end{itemize}
  \end{itemize}
\item {\bf API}
  \begin{itemize}
  \item {\bf Java} programs can be used to directly create both Generic and
    Geometric Models.
  \item {\bf C++} programs can be interfaced using proxies created by JACE~\cite{JACE}.
  \end{itemize}
\end{itemize}
Many of supported XML Schemas are supported also by other programs, like
Atlantis~\cite{Atlantis} or PersInt~\cite{PersInt}.

\section{Applications}

Various applications have been built on top of the GraXML Toolkit. The most
important (and most widely used) is the GraXML 3D Display.

\subsection{3D Display}

The GraXML Display offers the full functionality for browsing and manipulating 3D data. 
All the major User Requirements functionality for the 3D display are available:
\begin{itemize}
\item Standard 3D visual operations, like Rotation, Translation, Zooming, 
  Scaling, Sheering and Skewing. Those operations can be anizotropic, if
  applicable. All those operations can be applied either to the whole 
  Scene (Global Operations) or to the selected (picked) objects (Local 
  Operations). The Local Operation allows, for example, to move one object away 
  from others. Operations are available via a Drag mouse action, using a keyboard 
  or through the menu. The Go-to operation is also available, it moves the observer 
  closer to the selected object (or away from it).
\item Changing of visual properties of an object. The color of an object, its 
  transparency properties and its polygon status (solid - wireframe - 
  vertexframe)  can be changed interactively. An operation can be applied either 
  to the selected (picked) objects or the the whole SceneGraph.
\item Both Parallel and Perspective projections.
\item Cutting the 3D Scene from front and from back.
\item Selective switching of the objects on and off from the hierarchical 
  Tree Navigator. The switching works also in the optimized mode, the result 
  may be, however, surprising due to hidden relations.
\item Snapshot saving into the jpg file.
\item Context sensitive help and Context sensitive information about the Tree 
  Navigator elements.
\item Optimized or fully Interactive navigation. The Optimized mode uses much 
  smaller memory footprint and is therefore useful for the larger 3D Scenes 
  (more than tens of thousands of simple objects). Interactive features are 
  limited in this mode.
\item Quality level. Higher quality levels offer features like better 
  antialiasing or smoother objects, but requires significantly more CPU and 
  memory.
\item Output to VRML~\cite{VRML}. The VRML output can be also 
  translated into a POV~\cite{POV} Raytracing format file, which can then be used to 
  create photographic-quality pictures (incl. reflections, \dots).
\item Context sensitive introspection and actions allowing inspecting and 
  calling features of the real objects connected to the 3D objects. This 
  means that, for example, a user can re-fit a Track by interacting with its 
  graphical representation. This functionality, however, requires 
  a collaboration of a back-end (reconstruction) framework.
\item Full Java scripting interface. While complete Java environment is 
  available (including access to all GraXML objects), the intended use of the 
  scripting interface is to set various options, open files and clean the window.
  An example of a GraXML Display script is shown in the Listing~\ref{Script}.
\end{itemize}

\begin{figure*}[!]
\centering
\includegraphics[width=135mm]{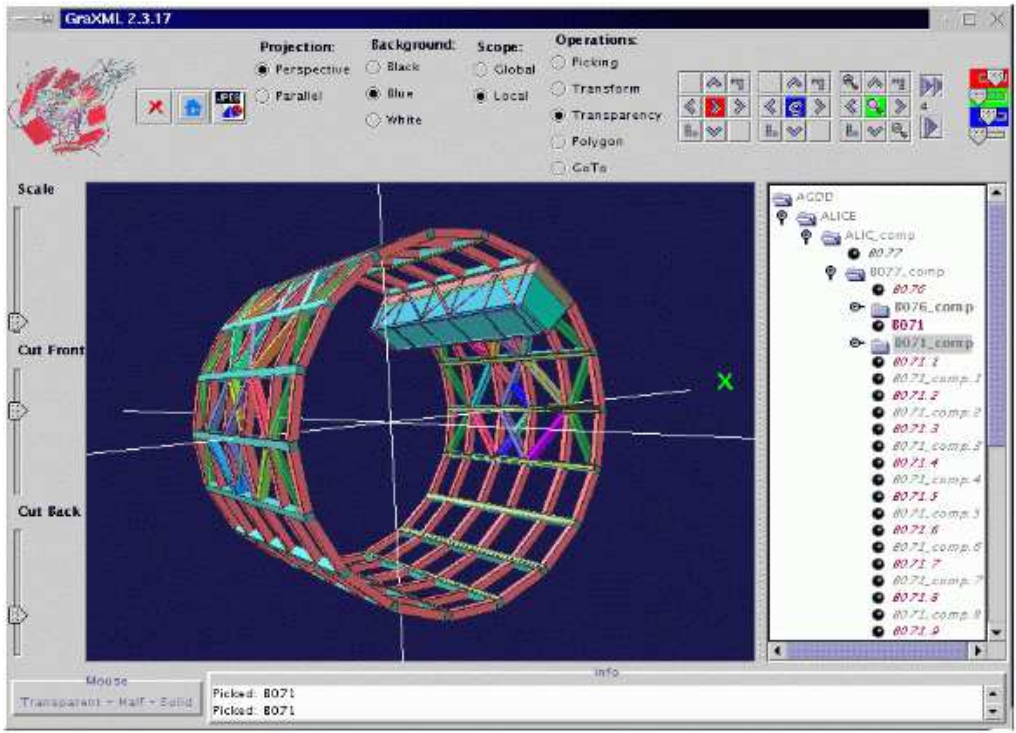}
\caption{ALICE Frame in the GraXML Display (the XML file has been generated from Geant4).} 
\label{ALICE}
\end{figure*}

\begin{figure*}[!]
\centering
\includegraphics[width=135mm]{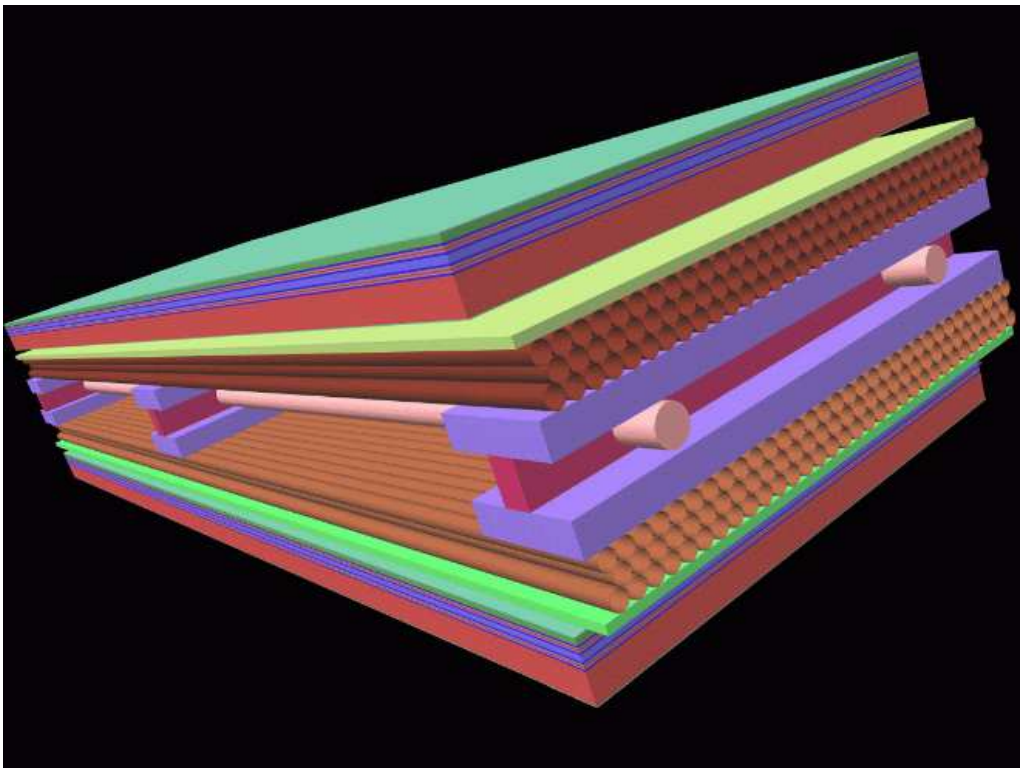}
\caption{ATLAS Muon Chamber.} 
\label{Chamber}
\end{figure*}

\begin{figure*}[!]
\centering
\includegraphics[width=135mm]{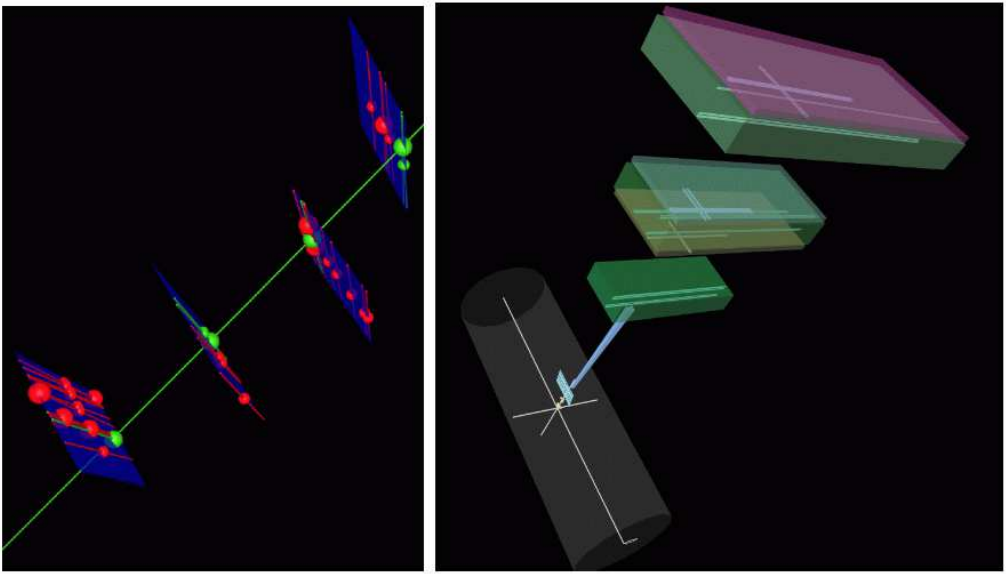}
\caption{Track in ATLAS Inner Detector and Muon Detector 
  - Photo-realistic View and Transparent Volumes.} 
\label{Track}
\end{figure*}

\begin{figure*}[!]
\centering
\includegraphics[width=135mm]{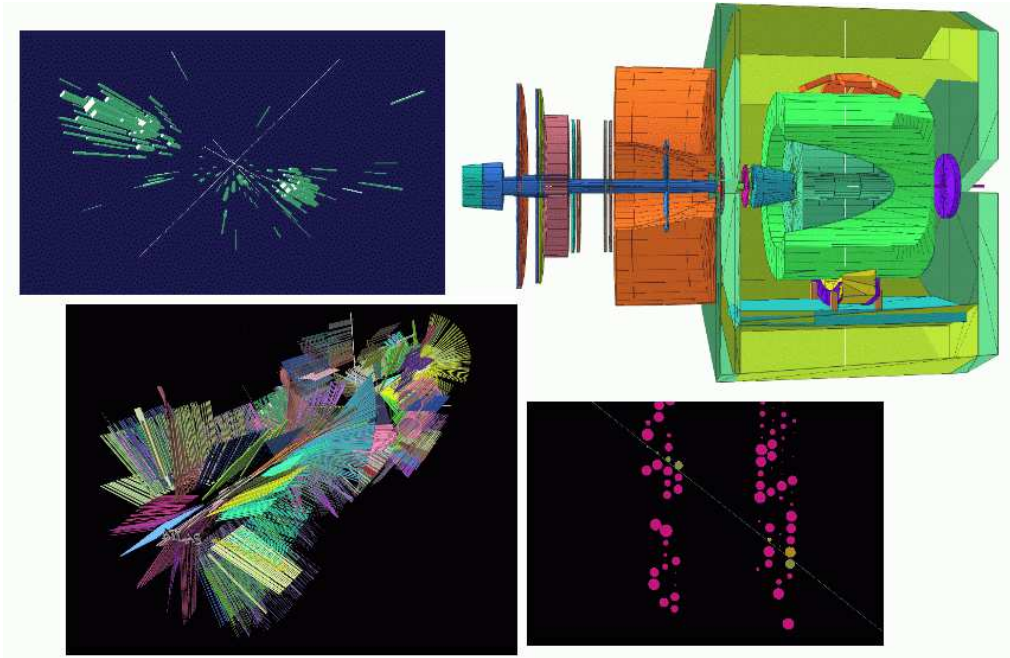}
\caption{1) ATLAS Event in Tile Calorimeter. 2) Semi-cutted, semi-transparent
  ALICE. 3) ATLAS Event in TRT. 4) ATLAS Track in Muon Chamber.} 
\label{Misc}
\end{figure*}

\begin{figure*}[!]
\centering
\includegraphics[width=135mm]{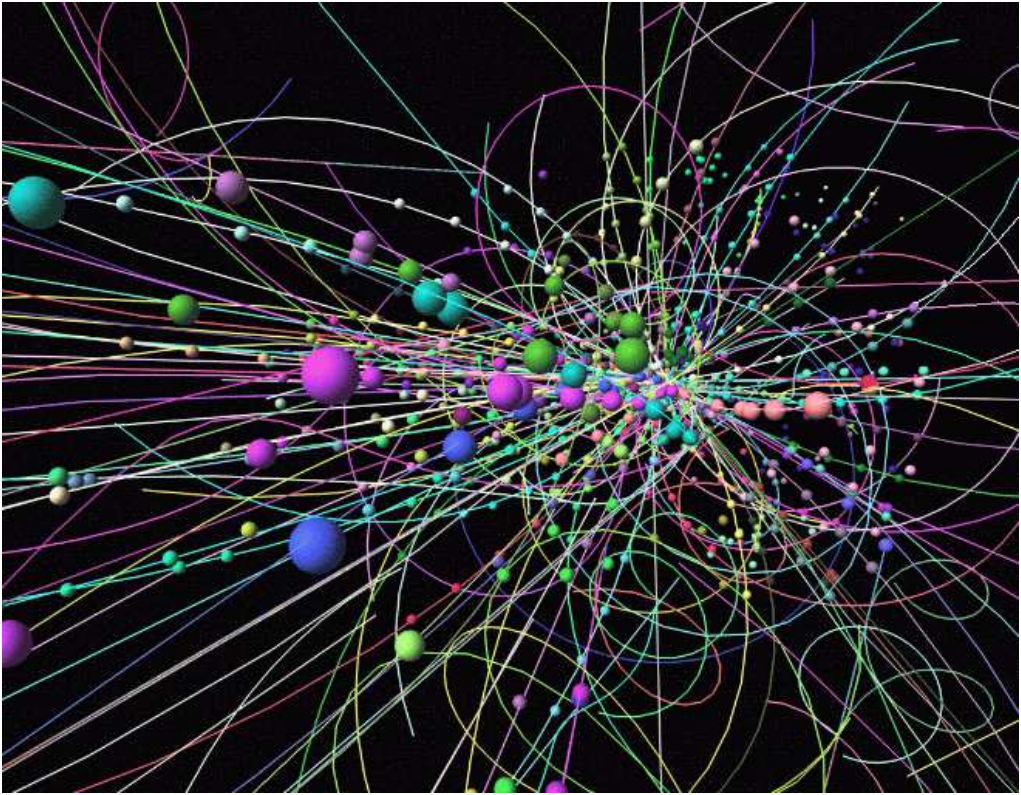}
\caption{Artistic view of the ATLAS Event in the Inner Detector.} 
\label{Art}
\end{figure*}

Several snapshots of running GraXML Display are shown in 
Figures~\ref{ALICE},~\ref{Chamber},~\ref{Track},~\ref{Misc},~\ref{Art}.

\begin{table}[!]
\begin{center}
\begin{tabular}{|l|}
\hline
$Configuration.setQuality(9);$ \\
$SelectedColor.setPalette(SelectedColor.ATLANTIS);$ \\
$TruthTrack.setPtCut(5.0);$ \\
$Hit.asSphere();$ \\
$Hit.colorFromKine();$ \\
$w.show("test.xml");$ \\
$j3d.snapshot("Picture.jpg");$ \\
\hline
\end{tabular}
\caption{Example of simple script for GraXML.}
\label{Script}
\end{center}
\end{table}

\subsection{Non-Graphical Applications}

The non-graphical Applications built on top of GraXML Toolkit are Exporters,
Converters and Importers:
\begin{itemize}
\item {\bf Exporters} for exporting geometric data into other Applications
  or file formats.
  \begin{itemize}
  \item Into VRML and X3D~\cite{X3D} format to be used by any VRML browser or in the 3D
    Cave.
  \item Into TXT format for easy debugging.
  \end{itemize}
\item {\bf Converters} for converting between various supported data formats.
  \begin{itemize}
  \item Between AGDD v4 and v6 via XSLT stylesheets.
  \item Between AtlasEvent and AtlantisEvent via a simple application.
  \end{itemize}
\item {\bf Importers} for importing from other application.
  \begin{itemize}
  \item Geant4 to AGDD via a simple C++ application provided in Virtual MC~\cite{VMC} built
    in ALICE.
  \end{itemize}
\end{itemize}

\section{Summary}

The GraXML toolkit provides a flexible foundation for modeling of 3D spacial data 
in HEP. Most of the required functionality is provided directly by Java3D, 
the rest is implemented on top. Applications built on GraXML can run in 
a Graphical or non-Graphical mode.

\newpage
\onecolumngrid

\end{document}